\documentstyle[11pt,fullpage,doublespace,testeq]{article}
\renewenvironment{thebibliography}[1] 
{\newpage\section*{References}\list
  {}{\leftmargin0.0in \labelwidth0.0in \labelsep0.0in
	\rightmargin0.0in \listparindent0.0in }
 \def\newblock{\hskip .11em plus .33em minus -.07em}
 \sloppy
 \sfcode`\.=1000\relax}{\endlist}
\newcommand{\p}{\partial}
\newcommand{\f}{\frac}
\newcommand{\eqref}[1]{(\ref{#1})}
\newcommand{\bel}[1]{\begin{equation}\label{#1}}
\newcommand{\be}{\begin{equation}}
\newcommand{\ee}{\end{equation}}
\newcommand{\bea}[1]{\begin{eqnarray}\label{#1}}
\newcommand{\ea}{\end{eqnarray}}
\newcommand{\bsa}[1]{\begin{speqa}\label{#1}}
\newcommand{\esa}{\end{speqa}}
\newcommand{\bsb}[1]{\begin{speqb}\label{#1}}
\newcommand{\esb}{\end{speqb}}

\newcommand{\wbf}{\mbox{\boldmath $\omega$}}
\newcommand{\twbf}{\mbox{\boldmath $\Omega_{T}$}}
\newcommand{\bwbf}{\mbox{\boldmath $\Omega$}}
\begin{document}
\begin{center}
{\Large {\bf An instability criterion for a finite amplitude 
localized \\*[0.1in] 
disturbance in a shear flow of electrically conducting fluids}} \\*[0.1in]
 by\\*[0.1 in] 
{{\large{\sc Vladimir Levinski, Ilia Rapoport and Jacob Cohen}}\\*[0.1 in]
{\footnotesize
 Faculty of Aerospace Engineering, Technion - Israel Institute of
Technology, Haifa 32000, Israel}}
\end{center}
\bigskip
\begin{center}
{\bf Abstract}
\end{center}

The stability of shear flows of electrically conducting fluids,
with respect to finite amplitude three-dimensional localized 
disturbances is considered. 
The time evolution of the fluid impulse integral, characterizing such 
disturbances, for the case of low magnetic Reynolds number 
is obtained by integrating analytically 
the vorticity equation.
Analysis of the resulted equation reveals a new instability criterion. 
\section{Introduction}
The use of magnetohydrodynamics (MHD) for transition and turbulent 
control is quite attractive from the point of view of applications, 
in particular when new technologies may allow 
direct turbulent control in sea water. 
Most of the work in this field has been concerned with flows of 
electrically conducting fluids subjected to transverse magnetic fields. 
This configuration is used in MHD generators, accelerators and 
pumps. Recently, Nosenchuck and Brown (1993), demonstrated 
experimentally that the application of wall-normal Lorentz force prohibit 
lift-up and bursting of near wall fluid, which are characteristics of 
the end-stage of transition and near wall turbulence. However, in such 
a configuration, the magnetic effects are due mainly to the coupling 
between the mean velocity profile and the magnetic field, rather than 
the damping of turbulence. 

On the other hand, when the mean flow and the magnetic field are aligned, 
the direct effect of the magnetic field is on the disturbed velocity field. 
Fraim and Heiser (1968) studied experimentally the effect of a strong 
longitudinal magnetic field on the  flow of mercury in a circular tube. 
They found that the magnetic field can significantly increase the 
critical Reynolds number ($Re_{cr}$) for transition. More examples showing 
similar results are summarized in the book of Branover (1978), 
who concluded that linear theories of stability of MHD flows yield 
values of $Re_{cr}$ which are much higher than the measured values. 
This suggests that a proper explanation of the above mentioned 
experimental results must take nonlinear effects into account. 

The evolution of a finite amplitude three-dimensional localized 
vortex disturbance 
embedded in an external incompressible shear flow was considered by 
Levinski \& Cohen (1995, hereinafter reffered to as LC).  
Using the fluid impulse as an integral characteristic of such a disturbance, 
they found that parallel shear flows are 
always unstable with respect to localized disturbances, 
the  typical dimension 
of which $\delta$, is much smaller than a dimensional length scale 
$\Delta$, corresponding to an ${\cal O}(1)$ change of the external velocity. 
Moreover, their analysis 
predicts that the growing vortex disturbance is inclined at
$45^{0}$ to the external flow direction, in a plane normal to the transverse
axis. It was also shown that although viscosity plays 
a crucial role in the generation of the initial localized disturbance 
and in determining the mean flow field, 
it plays no role in the time evolution of its fluid impulse integral. 
In other words, once the mean field is established, the subsequent 
evolution of the disturbance fluid impulse integral is largely an 
inviscid one. 
These predictions agree with previous experimental observations 
concerning the growth of hairpin vortices in laminar and turbulent 
boundary layers, see e.g. Head \& Bandyopadhyay (1981),   
Acarlar \& Smith (1987) and Hagen \& Kurosaka (1993).
The application of this approach to 
Taylor-Couette flow revealed a new instability criterion, which was 
recently verified experimentally  by Cohen {\it et al.} (1996).  

The purpose of the work reported here is to examine the effect of an 
externally imposed magnetic field on the onset of such disturbances. 
The analysis is restricted to incompressible shear flows 
characterized by a low magnetic Reynolds number, 
$Re_m = \mu \sigma \Delta U \ll 1,$ 
 where 
$\mu$ is the magnetic permeability, 
$\sigma$ is the electrical conductivity and  $U$ is a
characteristic velocity scale of the external flow. 
The magnetic Reynolds number can be considered as the 
ratio between the diffusion time scale of the magnetic field 
$ \mu \sigma \Delta^2$, 
and the hydrodynamic time scale 
$\Delta/U$. 
For small values of $Re_m$, 
the magnetic field induced by a disturbance diffuses rapidly. This 
leads to the dissipation of the disturbance 
kinetic energy and consequently may stabilize the flow. 
\section{Analysis}
For $Re_m \ll 1$ and stationary external 
magnetic field, it was shown by Braginskii (1960) that the 
electromagnetic force per unit volume is 
${\bf f} = \sigma\left[ ({\bf U}_T \times {\bf B}) \times
{\bf B} - {\bf \nabla} \Phi_T \times {\bf B} \right]$,
where ${\bf B}$ is the magnetic induction of the external field and 
${\bf U}_T$ is the total velocity vector as defined below. 
The scalar potential 
$\Phi_T$ is determined from the condition that the charge is neutralized, 
which for a uniform liquid yields 
\bel{quasi}
{\bf \nabla}^2 \Phi_T = {\bf B} \cdot {\bf \Omega}_T, 
\ee
where ${\bf \Omega}_T= {\bf \nabla} \times {\bf U}_T$ is the total vorticity
vector. For this case, 
the three-dimensional vorticity equation for an incompressible 
flow is given by
\bel{maxv}
\f{\p {\bf \Omega}_T}{\p t} + ({\bf U}_T \cdot {\bf \nabla}){\bf \Omega}_T
- ({\bf \Omega}_T \cdot {\bf \nabla}){\bf U}_T - \f {\sigma}{\rho } 
({\bf B} \cdot {\bf \nabla}) \left( {\bf U}_T \times
{\bf B} - {\bf \nabla} \Phi_T \right)  = \nu \Delta {\bf \Omega}_T, 
\ee
where $\rho$ and $\nu$ are the density and kinematic viscosity of the 
fluid, respectively. 

We consider the flow field as being the sum of two contributions: 
the external shear field in which 
$\bwbf= {\bf \nabla} \times {\bf U}$, 
and a finite amplitude disturbed field in which  
$\wbf= {\bf \nabla} \times {\bf u}$. Thus, the total velocity and vorticity 
vectors can be written as ${\bf U}_T ={\bf U} +{\bf u}$ and 
$\twbf=\bwbf+\wbf$, 
where the undisturbed external flow field is assumed to be a known solution
of \eqref{quasi} and  \eqref{maxv} when ${\bf u}=0$. 
The initial disturbed vorticity 
${\wbf({\bf x},t_0)}={\wbf}_0({\bf x})$ is 
assumed to be confined to the 
small region of order $\delta \ll \Delta$ as well as $\delta \ll\Delta_1$, 
where $\Delta_1$ is a typical length 
scale corresponding to an ${\cal O}(1)$ change of the external 
magnetic field. 

Owing to the smallness of the 
disturbed region, the external velocity, vorticity and 
magnetic fields are approximated by Taylor series expansions. 
Following LC, we use a Galilean frame, moving with the 
disturbance, i.e., ${\bf U}(0)=0$, and  consider 
the initially embedded vorticity region as surrounded by 
an infinite field having a constant velocity shear and a constant magnetic 
induction. Consequently, 
\bel{tay}
U_i({\bf x}) = \sum_{j=1}^{3}\f{\p U_i(0)}{\p x_j}x_j, \;\;\;\;
 \Omega_i({\bf x}) = \Omega_i({0}),\;\;\;\;
B_i({\bf x}) = B_i({0}) \;\;\;\; where \;\;\;i=1,2,3. 
\ee
When the equations for the undisturbed external 
flow and magnetic field are subtracted from \eqref{quasi} and 
\eqref{maxv} respectively, we obtain
\bel{euler}
\f{\p \wbf}{\p t} + 
({\bf U \cdot \nabla})\wbf - (\wbf {\bf \cdot \nabla}){\bf U} 
- ({\bf \Omega \cdot \nabla}){\bf u } + ({\bf u \cdot \nabla})\wbf 
- (\wbf{\bf \cdot \nabla}) {\bf {u}} 
- \f {\sigma}{\rho } 
({\bf B} \cdot {\bf \nabla}) \left( {\bf u} \times
{\bf B} - {\bf \nabla} \phi \right)  = 0, 
\ee
and 
\bel{vorticity}
 {\bf \nabla}^2 \phi ={\bf B} \cdot {\bf \omega}. 
\ee
In \eqref{euler} the viscous term is omitted since, as was shown by 
LC, 
the corresponding contribution of the viscous term to the dynamics 
of the fluid impulse integral 
vanishes in view of the asymptotic behavior of the 
disturbance vorticity far from the origin. 

We shall follow the evolution in time of the fluid impulse integral of
the disturbance, defined as
\bel{lamb}
 \f{d {\bf P}}{dt} = \frac{1}{2} \int {\bf x}\times
   \f{\p \wbf({\bf x},t)}{\p t}{\it d V},
\ee
where  ${\bf x}$ is the 
position vector,  ${\it dV}$ is a
volume element and 
the time derivative of $\wbf({\bf x},t)$ is determined from 
\eqref{euler}.

Since the time evolution of the fluid impulse is 
an integral over the whole volume, we must first verify that 
most of the contribution to this integral comes from the localized 
disturbed region. 
Indeed, all of the vorticity contributing to this integral, 
except for the part generated via the fourth 
and the seventh (electromagnetic) terms in \eqref{euler}, is confined 
to the disturbed region. 

In order to estimate the contributions of the fourth and the 
seventh 
terms in \eqref{euler}, we examine their far-field behavior. 
The expression for the scalar potential $\phi$ is obtained from 
the general solution of \eqref{vorticity} 
\bel{phi}
\phi = \f{1}{4\pi} \int \f{{\bf B} \cdot {\bf \omega}({\bf x}_1)}
{|{\bf x}-{\bf x}_1|^3} {\it  dV}_1 = {\bf B} \cdot {\bf M},
\ee
where the asymptotic series of ${\bf M}$, expressed 
in terms of the fluid impulse, is given (Batchelor, 1967) by 
\bel{asym}
{\bf M} = \f{1}{4\pi} {\bf P} \times \f {{\bf x}}{|{\bf x}|^3} +
{\cal O}(\f {1}{|{\bf x}|^3}). 
\ee
Similarly, the far-field velocity induced by the localized 
vortex disturbance is 
\bel{asymp}
{\bf u}({\bf x}) \sim -\frac{1}{4\pi}\left[\frac{{\bf P}}{\left|
{\bf x}\right|^3} - \frac{3({\bf P} \cdot {\bf x}) 
 {\bf x}}{\left| {\bf x}\right|^5}\right] 
+ {\cal O}\left(\frac{1}{\left| {\bf x}\right|^4}\right). 
\ee
Substitution of \eqref{asym} and \eqref{asymp} into \eqref{euler} 
shows that the far-field vorticity diminishes in magnitude as 
$|{\bf x}|^{-4}$. Consequently, the integral \eqref{lamb} is not absolutely 
convergent and depends on the way in which the volume of integration 
is allowed to tend to infinity. 

In order to overcome this difficulty we follow the procedure proposed
in LC. Accordingly, we subdivide the velocity and 
vorticity fields into two 
parts, $\wbf={\wbf}^{I}+{\wbf}^{II}$ and 
${\bf u } ={\bf u}^{I} +{\bf u}^{II}$,  
so that ${\wbf}^{I,II}={\bf \nabla}\times {\bf u}^{I,II}$. 
Therefore, for each part we require that
\bel{div}
{\bf \nabla}\cdot  \wbf^{I}={\bf \nabla}\cdot \wbf^{II} =0. 
\ee
The first part, indicated by the superscript $I$,  is 
associated with the concentrated 
vorticity confined within and in the vicinity of the initially disturbed 
region, whereas  the second part, indicated by the superscript $II$, 
is associated with the far-field vorticity 
generated by the problematic terms (the fourth and seventh) 
in \eqref{euler}. 
Accordingly,  we set the initial distribution of the vorticity fields as:
\bel{sub0}
\wbf^{I}({\bf x},t=t_0)=\wbf_0({\bf x}) \;\;\;\;\;\; and \;\;\;\;\;\;
\wbf^{II}({\bf x},t=t_0)=0, 
\ee
and follow the evolution of $\wbf^{I}({\bf x},t)$. 
In addition, we subdivide the whole space into two regions, inside 
and outside a spherical domain of radius 
$R \geq \delta$, enclosing the disturbance. In the outer region the 
corresponding system of the vorticity equations is given by 
\begin{displaymath}
\f{\p \wbf^{I}}{\p t} + ({\bf U \cdot \nabla})\wbf^{I} - (\wbf^{I} 
{\bf \cdot \nabla}){\bf U} + ({\bf u \cdot \nabla})\wbf - 
(\wbf{\bf \cdot \nabla}) {\bf u} - ({\bf \Omega \cdot \nabla})
({\bf u}^{I} - {\bf u}_0)  
\end{displaymath}
\bel{subwI}
- \f {\sigma}{\rho } ({\bf B} \cdot {\bf \nabla}) \left[ 
({\bf u}^{I}-{\bf u}_0) \times {\bf B} - {\bf \nabla} \left({\bf B} \cdot
({\bf M}^I - {\bf M}_0) \right) \right] = 0, 
\ee
\begin{displaymath}
\f{\p \wbf^{II}}{\p t} + ({\bf U \cdot \nabla})\wbf^{II} 
- (\wbf^{II} {\bf \cdot \nabla}){\bf U} - ({\bf \Omega \cdot \nabla})
({\bf u}^{II} + {\bf u}_0) 
\end{displaymath}
\bel{subwII}
- \f {\sigma}{\rho } ({\bf B} \cdot {\bf \nabla}) \left[ 
({\bf u}^{II}+{\bf u}_0) \times {\bf B} - {\bf \nabla} \left({\bf B} \cdot
({\bf M}^{II} + {\bf M}_0) \right) \right] = 0, 
\ee
where ${\bf u}_0$ and ${\bf M}_0$ are the leading terms in the asymptotic 
series of \eqref{asym} and \eqref{asymp}, 
\bel{des}
{\bf u}_0 = - \f{1}{4\pi} \left[\f{{\bf p}}{|{\bf x}|^3}  -
\f{3 ({\bf p} \cdot {\bf x}) {\bf x}}{|{\bf x}|^5} \right] \;\;\; ;
\;\;\; {\bf M}_0 = \f{1}{4\pi} {\bf p} \times \f {{\bf x}}{|{\bf x}|^3}, 
\ee
and the fluid impulse ${\bf p}$ corresponds only to $\wbf^{I}$, i.e.,
\bel{lambI}
    {\bf p} = \frac{1}{2} \int {\bf x}\times
    \wbf^{I}({\bf x}){\it dV}. 
\ee

Since ${\bf u}_0$ and ${\bf M}_0$ cancel the leading terms of the 
far-field vorticity, generated via the problematic terms in \eqref{subwI}, 
the asymptotic behavior of 
$\wbf^{I}$ is given by
\bel{wasym}
     \left|\wbf^{I}({\bf x},t)\right| \sim {\cal O}\left( \f{1}{\left|
     {\bf x}\right|^5}\right) \;\; for \;\; \left|{\bf x}\right| 
\gg \delta, 
\ee
and therefore the  fluid impulse integral \eqref{lambI} is absolutely 
convergent. 

For the inner region we write 
\begin{displaymath}
\f{\p \wbf^{I}}{\p t} + ({\bf U \cdot \nabla})\wbf^{I} - (\wbf^{I} 
{\bf \cdot \nabla}){\bf U} + ({\bf u \cdot \nabla})\wbf - 
(\wbf{\bf \cdot \nabla}) {\bf u} - ({\bf \Omega \cdot \nabla})
{\bf u}^{I}  
\end{displaymath}
\bel{subwIn}
- \f {\sigma}{\rho } ({\bf B} \cdot {\bf \nabla}) \left( 
{\bf u}^{I} \times {\bf B} - {\bf \nabla} ({\bf B} \cdot
{\bf M}^I ) \right) + {\bf \nabla} \Psi = 0, 
\ee
\bel{subwIIn}
\f{\p \wbf^{II}}{\p t} + ({\bf U \cdot \nabla})\wbf^{II} 
- (\wbf^{II} {\bf \cdot \nabla}){\bf U} - ({\bf \Omega \cdot \nabla})
{\bf u}^{II} - \f {\sigma}{\rho } ({\bf B} \cdot {\bf \nabla}) \left( 
{\bf u}^{II} \times {\bf B} - {\bf \nabla} ({\bf B} \cdot
{\bf M}^{II} ) \right) - {\bf \nabla} \Psi = 0, 
\ee
so that the sum of the two equations in each region yields Eq. \eqref{maxv}  
for that region, and together with the 
initial conditions given in \eqref{sub0}, yields the original 
problem for the entire space. 

For the outer region, condition \eqref{div} is always satisfied. 
For this condition to be satisfied in the inner region 
${\bf \nabla}^2\Psi$ must be equal to zero, as can be shown by 
applying the
operator $({\bf \nabla \cdot})$ to \eqref{subwIn} and \eqref{subwIIn}. 
Then, $\Psi$ is determined by solving the 
Neumann problem for which the normal derivative of $\Psi$ at $|{\bf x}|=R$ 
is matched with the scaler product of the unit vector normal to the 
boundary surface ${\bf n}$,  
and the terms in \eqref{subwI}, containing ${\bf u}_0$ and ${\bf M}_0$,  
i.e. 
\bel{bound}
{\bf n} \cdot \f {\p \Psi}{\p {\bf n}} 
  {\left |\vbox to 17pt {} \right.}_{|{\bf x}|=R} 
= {\bf n} \cdot \left[ ({\bf \Omega \cdot \nabla}){\bf u}_0 +
\f {\sigma}{\rho } ({\bf B} \cdot {\bf \nabla}) \left({\bf u}_0 
\times {\bf B} - {\bf \nabla} ({\bf B} \cdot {\bf M}_0)\right) 
\right]
{\left |\vbox to 17pt {} \right.}_{|{\bf x}|=R}. 
\ee
%
Accordingly, using \eqref{des} the expression for $\Psi$ is given by
\bel{psib}
\Psi = \f {3}{8 \pi R^4} \left[({\bf \Omega} \cdot {\bf p})|{\bf x}|^2 - 
3 ({\bf \Omega} \cdot {\bf x})({\bf p} \cdot {\bf x}) + 
\f {4 \sigma}{\rho } ({\bf B} \cdot {\bf x})
({\bf x} \cdot ({\bf p} \times {\bf B})) \right].
\ee
Since the integral \eqref{lambI} is absolutely 
convergent, it is convenient to use an infinite sphere as the volume of 
integration. Consequently, the time evolution of ${\bf p}$ is given by 
\bel{dpdtI}
 \f{d {\bf p}}{dt} = \frac{1}{2} \lim_{R_1 \rightarrow \infty}
 \int_{\left|{\bf x}\right|\leq R_1} {\bf x}\times
   \f{\p \wbf^{I}({\bf x},t)}{\p t}{\it d V}. 
\ee
Substitution of \eqref{subwI} and \eqref{subwIn} into \eqref{dpdtI} yields
\begin{displaymath}
 \f{d {\bf p}}{dt} = - \frac{1}{2} \lim_{R_1 \rightarrow \infty} 
 \int_{\left|{\bf x}\right|\leq R_1} {\bf x}
 \times \left[({\bf U \cdot \nabla})\wbf^{I} - (\wbf^{I} 
 {\bf \cdot \nabla}){\bf U}-({\bf \Omega \cdot \nabla})
 {\bf u^{I}} + ({\bf u \cdot \nabla})\wbf - (\wbf{\bf \cdot \nabla}) 
 {\bf u} \right]{\it d V} 
\end{displaymath}
\begin{displaymath}
+ \f{\sigma}{2 \rho } \lim_{R_1 \rightarrow \infty} 
 \int_{\left|{\bf x}\right|\leq R_1} {\bf x}
 \times \left[ ({\bf B} \cdot {\bf \nabla}) 
 \left({\bf u}^{I} \times {\bf B} - {\bf \nabla} 
 ({\bf B} \cdot {\bf M}^I ) \right) \right]{\it d V} - \frac{1}{2} 
 \lim_{R_1 \rightarrow \infty} \int_{R \leq \left|{\bf x}\right|
 \leq R_1} {\bf x} \times \left[({\bf \Omega \cdot \nabla}) {\bf u_{0}}  
 \right.
\end{displaymath}
\bel{subst}
 \left. + \f {\sigma}{\rho } ({\bf B} \cdot {\bf \nabla}) {\bf u}_{0}
 \times {\bf B}\right] {\it d V} +  \f {\sigma}{2 \rho } \lim_{R_1 
 \rightarrow \infty} \int_{R \leq \left|{\bf x}\right| \leq R_1} 
 {\bf x} \times ({\bf B} \cdot {\bf \nabla}){\bf \nabla}({\bf B} \cdot 
 {\bf M}_0) {\it d V} - \frac{1}{2} \int_{\left|{\bf x}\right| \leq R} 
 {\bf x} \times {\bf \nabla} \Psi {\it d V}.
\ee
Each one of the integrals in \eqref{subst} is evaluated 
for a finite value of $R_1$ and its limit as 
$R_1 \rightarrow \infty$ is then taken. 
The result of the first integral in \eqref{subst} was already obtained 
in LC. As  shown in the Appendix for such domains of 
integration, 
the integral contribution of the last three integrals in \eqref{subst} 
is identically zero. Therefore, the artificial vorticity field  
has no direct impact on the evolution of the fluid impulse 
associated with the concentrated 
vorticity $\wbf^{I}({\bf x},t)$. Moreover, as was shown in 
LC, the influence of 
the vorticity field $\wbf^{II}({\bf x},t)$ on the evolution of 
$\wbf^{I}({\bf x},t)$ can be neglected. 
The second integral in \eqref{subst} is 
calculated using a similar procedure to that described in the Appendix 
and in LC. 
Finally, \eqref{subst} becomes 
\bel{befin}
 \f{d {\bf p}}{dt} = - \frac{1}{2} {\bf \nabla}({\bf p} \cdot {\bf U})-
 \frac{1}{2} ({\bf p} \cdot {\bf \nabla}){\bf U} -
 \f {2 \sigma B^2}{5 \rho } {\bf p} +
\f {\sigma }{5 \rho } ({\bf B} \cdot {\bf p}) {\bf B}. 
\ee
\section{Application to a representative example}
In the following we consider a simple example in which \eqref{befin} 
is applied to a parallel shear flow of electrically conducting fluids 
and a new 
instability criterion for finite amplitude localized disturbances is found. 
For a parallel plane shear flow,  the external 
velocity field is given by ${\bf U}=\left(U(y),0,0\right)$ , for which 
a right-handed coordinate system is used with ${\bf x}=(x,y,z)$, where 
the vector entries are the downstream, cross-flow and 
spanwise  directions, respectively.
The direction of the magnetic 
induction vector is chosen to be parallel to the downstream direction, 
i.e. ${\bf B}=\left(B,0,0\right)$. 
As was mentioned above, such a flow does not exhibit a direct coupling
between the mean flow and the magnetic field. Consequently, 
the direct effect of the magnetic field on the turbulent structure 
can be revealed. 
In this case, 
equation \eqref{befin} for the fluid impulse vector 
${\bf p}=(p_x,p_y,p_z)$, is reduced to 
\bel{evo1}
   \f {d p_x}{d t} = - \frac{1}{2} p_y \frac{d U}{d y}
   - \f {\sigma B^2}{5 \rho } p_x, \;\;\;\;
   \f {d p_y}{d t} = - \frac{1}{2} p_x \frac{d U}{d y}
   - \f {2 \sigma B^2}{5 \rho } p_y, \;\;\;\;
   \f {d p_z}{d t} = - \f {2 \sigma B^2}{5 \rho } p_z, 
\ee
for which the eigenvalues $\{\lambda_i\}_{i=1}^{3}$ can be found from 
the characteristic equation 
\bel{rate}
  \left(\lambda_i+ \f {2 \sigma B^2}{5 \rho } \right)
  \left[\lambda_i^2+ \f {3 \sigma B^2}{5 \rho } \lambda_i +
  2 \left(\f {\sigma B^2}{5 \rho } \right)^2 - 
  \left(\frac{1}{2} \frac{d U}{d y} \right)^2 \right] = 0.
\ee
Hence, the flow under investigation is stable with respect to 
three-dimensional localized disturbances only if the real part of 
$\lambda_i$ is not positive. Therefore, for stability we require that
\bel{uneq}
  2 \left(\f {\sigma B^2}{5 \rho } \right)^2 \geq
  \left(\frac{1}{2} \frac{d U}{d y} \right)^2  \;\;\; or \;\;\; 
  N=\f {\sigma B^2}{\rho  dU/dy} \geq \f {5}{2 \sqrt{2}} \; ,
\ee
where $N$ is a dimensionless interaction parameter which represents the ratio 
between the electromagnetic and the inertia forces. 

The term `stability' (and `instability') here 
means that the fluid impulse of a closed localized (in all three directions) 
vortex disturbance  will not 
grow (or grow) in time. The fluid impulse is a very suitable characteristic 
of localized vortex structures such as hairpin vortices in boundary layers, 
since it combines the geometrical dimensions of the structure together 
with the magnitude of its vorticity field. 
As such, this stability definition cannot describe `wavy' disturbances or 
`quasi' two-dimensional structures for which the fluid impulse integral 
is not defined. 
It should be noted however, that the above stability definition is not
equivalent to the conventional criteria of linear stability and energy 
stability. In fact, the growth of the fluid impulse 
does not necessarily guaranty growth in energy of the localized 
disturbance. For example, viscous diffusion leads 
to the decay of the localized disturbance energy while its fluid impulse 
remains the same. 

As an example of the new stability criterion, the value of the interaction 
parameter required 
for stability of Poiseuille flow in a tube is 
$ \sigma B^2 D / \rho  \bar{U} \geq 14.1$, where 
$D$ is the tube diameter and $\bar{U}$ is the mean velocity. 
For flow of mercury in a circular tube subjected to a strong 
longitudinal magnetic field, Fraim \& Heiser  
reported an increase of the critical Reynolds number for transition 
from 2250 to 10350 when the 
magnetic induction $B$ was increased from zero to 
$1.75 weber/m^2$,  and the corresponding interaction parameter 
at the upper limit was 9. As can be seen in Fig.~10 of their article, 
this value of the interaction parameter is 
an order of magnitude larger than the value predicted by Stuart's 
linear theory (Stuart, 1954). A  direct comparison of these 
results with our prediction 
is questionable because of at least two reasons. The first is that the 
experimental mean velocity profiles were not reported by 
Fraim \& Heiser. The second is  that the criterion presented 
here is applicable only with respect to localized disturbances 
whereas transitional 
flows include various types of disturbances. Nevertheless, the predicted 
value of the interaction parameter found in this paper is of the same order of 
magnitude as that of the experimental ones for large Reynolds numbers.  

\bigskip
\centerline{\bf ACKNOWLEDGMENTS}
\bigskip
The authors are grateful to J. Tanny for his careful reading of 
the manuscript and helpful discussion. 
The research was supported by Grant No. 3875-2-93 from the Israeli 
Ministry of Science and Technology. 

\section*{Appendix }
\vspace{0.15in}
\setcounter{equation}{0}
\setcounter{appeqno}{0}
\addtocounter{appsect}{1}
In this Appendix we show that the last three integrals of \eqref{subst}, 
defined as ${\bf  J}^1$, ${\bf J}^2$ and ${\bf J}^3$, respectively, 
are identically zero. 
In Cartesian tensor notation, the sum of ${\bf J}^2$ and ${\bf J}^3$ is 
\begin{appeqa}
J^2_i + J^3_i = \f {\sigma}{2 \rho } 
\epsilon_{ijk} \int_{R \leq \left|{\bf x}\right| \leq R_1} 
x_j \frac {\partial }{\partial x_{k}} ({\bf B} \cdot {\bf \nabla})
({\bf B} \cdot {\bf M}_0) {\it d V} - \frac{1}{2} \epsilon_{ijk}
\int_{\left|{\bf x}\right| \leq R} x_j \frac {\partial \Psi}
{\partial x_{k}}{\it d V},
\end{appeqa}
where $\epsilon_{ijk}$ is the alternating tensor and the usual 
summation convention is applied. 

Integration by parts and using Gauss' divergence theorem 
yields 
\begin{displaymath}
J^2_i + J^3_i = \f {\sigma}{2 \rho } 
\epsilon_{ijk} \left[ \oint_{\left|{\bf x}\right|=R_1} n_k x_j
({\bf B} \cdot {\bf \nabla})({\bf B} \cdot {\bf M}_0) dS -
\oint_{\left|{\bf x}\right|=R} n_k x_j
({\bf B} \cdot {\bf \nabla})({\bf B} \cdot {\bf M}_0) dS \right. 
\end{displaymath}
\begin{appeqa}
\left. - \delta_{jk} \int_{R \leq \left|{\bf x}\right| \leq R_1} 
({\bf B} \cdot {\bf \nabla})({\bf B} \cdot {\bf M}_0) {\it d V} 
\right] - \frac{1}{2} \epsilon_{ijk} \left[\oint_{\left|{\bf x}\right|=R}
n_k x_j \Psi dS - \delta_{jk} \int_{\left|{\bf x}\right| \leq R} 
\Psi {\it d V} \right], 
\end{appeqa}
where $\delta_{ij}$ is the Kronecker delta function. 
Using the properties of the alternating and symmetrical tensors 
$ \epsilon_{ijk} \delta_{jk} \equiv 0$, and 
$ \epsilon_{ijk} n_k n_j \equiv 0$, it follows that 
${\bf J}^2 = {\bf J}^3 = 0$. 

A similar procedure for ${\bf J}^1$ gives 
\begin{displaymath}
{\bf J}^1 = \frac{1}{2} \left(\oint_{\left|{\bf x}\right|=R_1} -
\oint_{\left|{\bf x}\right|=R} \right) \left[d{\bf S}({\bf \Omega} \cdot 
{\bf M}_0) - (d{\bf S} \cdot {\bf \Omega}) ({\bf M}_0 + {\bf x} 
\times {\bf u}_{0}) \right.
\end{displaymath}
\begin{appeqa}
\left. + \f {\sigma}{\rho } \left[B^2 d{\bf S} \times {\bf M}_0 -
{\bf B} ((d{\bf S} \times {\bf M}_0) \cdot {\bf B}) -
(d{\bf S} \cdot {\bf B})({\bf x} \times ({\bf u}_{0} \times {\bf B}))\right]
\right].   
\end{appeqa}
When \eqref{des} is substituted into (A3), the result of the two 
surface integrals become independent of the surface radii and consequently 
cancel each other, i.e., ${\bf J}^1=0$.

\newpage
\singlespace 
\end{document}